\documentclass[11pt]{article}

\usepackage[T1]{fontenc}
\usepackage[utf8]{inputenc}
\usepackage[margin=1in]{geometry}

\usepackage{amsmath,amssymb}
\usepackage{booktabs}
\usepackage{csquotes}
\usepackage{graphicx}
\usepackage{tikz}
\usepackage{xcolor}
\usetikzlibrary{arrows.meta,positioning,decorations.pathreplacing}

\usepackage[numbers,sort&compress]{natbib}
\usepackage[hidelinks]{hyperref}

\newcommand{\rev}[1]{#1}
\providecommand{\Description}[1]{}

\newcommand\blfootnote[1]{%
  \begingroup
  \renewcommand\thefootnote{}\footnote{#1}%
  \addtocounter{footnote}{-1}%
  \endgroup
}

\title{Frankenstein in the Pipeline:\\ Computational Epistemicide in Facial Recognition}

\author{%
  Nina da Hora\\[2pt]
  \normalsize Universidade Estadual de Campinas, Campinas, S\~ao Paulo, Brazil\\
  \normalsize Instituto da Hora, Brazil\\
  \normalsize \texttt{ninadhoraa@gmail.com}
}
\date{}

\begin{document}
\maketitle

\blfootnote{Author's version of a paper accepted to the 2026 ACM Conference on Fairness, Accountability, and Transparency (FAccT~'26), June 25--28, 2026, Montreal, QC, Canada. The definitive Version of Record is published in the conference proceedings: \href{https://doi.org/10.1145/3805689.3812284}{https://doi.org/10.1145/3805689.3812284}. \textcopyright{} 2026 Nina da Hora. Published by ACM under a Creative Commons Attribution 4.0 International (CC BY 4.0) License.}

\begin{abstract}
While the eugenic roots of computer vision are well-documented in critical technology studies, less attention has been paid to the operational mechanisms through which this violence is enacted at the level of the pipeline. This paper employs Mary Shelley's \textit{Frankenstein} not as a metaphor for unintended consequences, but as a diagnostic framework for method: disassembly, reconstruction, and the production of a creature whose legitimacy is asserted by the procedure that made it.

I argue that embedding-based facial recognition enacts what I call \emph{computational epistemicide}, an extension of Sueli Carneiro's concept of epistemicide to the computational domain---by destroying the face as a living, relational surface and authorizing a numerical proxy as the privileged site of identity. Across detection/cropping, landmarking, alignment/frontalization, and embedding, the face is progressively narrowed to what can be stabilized as data, producing a \emph{canonical face} as the condition of legibility and a corresponding \emph{form-subject} as the condition of recognition. Vectorization completes the Frankensteinian ``stitching'': the dissected face is reassembled into a fixed-dimensional artifact designed to circulate across databases and institutions. I then show how distance-based similarity and thresholding operationalize a norm of ``close enough,'' making recognition inseparable from standardization and rendering reformist ``ethical AI'' optimization structurally insufficient. The paper concludes by arguing for abolition as a normative stance: refusing vectorized identity as a legitimate basis for rights and access, and dismantling the institutional impulse to govern human life through dissectible data points.
\end{abstract}


\bigskip

\section{Introduction}
Facial recognition technologies are increasingly deployed as instruments of governance in public security, access control, and institutional verification under the promise that identity can be reliably inferred from facial images. The dominant critical diagnosis has emphasized data bias and disparate error rates, showing how performance disparities track race and gender and how these systems reproduce inequality in practice \cite{buolamwini2018gender,grother2019frvt}. This paper agrees that these disparities matter, but argues that focusing only on performance risks missing a more basic source of harm.

I argue that the core violence of facial recognition is operational and ontological: contemporary pipelines transform the face into a fixed-dimensional proxy and then treat proximity in a learned geometry as identity. This is not a neutral ``encoding'' of the face; it is a regime of recognizability that strips the face of its relational, historical, and situated meanings in order to make it computable and governable. I use the concept of \emph{computational epistemicide}, first proposed by da Hora \cite{dahora2025contra,dahora2026imagens}, to describe this operational logic: the destruction of the face as a living relational surface through the epistemic operations required for recognition to function.

\rev{I write as a computer scientist trained to optimize and formalize, but increasingly situated at the borders between computational science and critical epistemology. The dominant critical response to facial recognition has focused on outcomes: biased datasets, disparate error rates, discriminatory deployments. That work is necessary and this paper builds on it. But the argument developed here operates at a different level: the philosophy of decision embedded in the pipeline itself. The architectural choices, the loss functions, the geometric constraints, the threshold logic are decisions made before any data is collected. They determine what counts as a face, as similarity, as identity. The epistemic violence is not downstream of these decisions; it is constituted by them. This paper opens a complementary front: the critical epistemology of the computational method itself, analyzed from within the technical apparatus.}

My use of \emph{epistemicide} extends the concept as developed by Sueli Carneiro: the production of the Black subject as ``non-being'' through systematic forms of erasure, negation, and the denial of epistemic authority \cite{carneiro2005construcao}. The computational extension \cite{dahora2025contra,dahora2026imagens}, developed in the context of algorithmic image regimes in Brazil, specifies how this denial is operationalized through pipelines that convert living faces into governable proxies. This paper operationalizes that concept across the face recognition pipeline and develops the Frankenstein diagnostic as its methodological framework. In face recognition, erasure is enacted through computation: the face is made knowable only by being reduced to what the system can stabilize as data, and what exceeds that grammar is rendered noise, distortion, or irrelevance. The result is not merely misrecognition but an epistemic displacement in which the vector becomes the authorized site of identity and the living face is subordinated to its proxy. Situated within the global production of race \cite{silva2007toward}, this displacement is not accidental: it inherits a historical order in which recognizability is organized by raciality and secured through surveillance \cite{browne2015dark}.

To make this argument, the paper uses \textit{Frankenstein} as a diagnostic framework rather than a cautionary metaphor: the point is not that technology has unintended consequences, but that a method of disassembly and stitching is constitutive of what embedding-based recognition is \cite{shelley1818frankenstein}. I connect this diagnostic to abolitionist critiques of racial technologies, in which the promise of neutral innovation often functions as a renewed regime of classification and control \cite{benjamin2019race}. Under this premise, reformist ``ethical AI'' optimization cannot resolve the harm by improving accuracy or calibration, because the harm is enacted by the representational method itself (Sections~\ref{sec:epistemicide}--\ref{sec:metric-politics} and Section~\ref{sec:reform-fails})

\paragraph{Contributions.}
This paper makes two contributions. First, it introduces and operationalizes the concept of \emph{computational epistemicide} in face recognition: a sequence of epistemic operations that destroys the face as a living relational surface and replaces it with a governable proxy. Second, it models the politics of similarity by showing how identity-as-distance and thresholding operationalize a norm of ``close enough,'' clarifying why reformist ethical optimization cannot resolve harms rooted in ontological substitution.

\paragraph{Roadmap.}
Section~\ref{sec:from-face-to-vector} establishes the minimal technical background and the embedding paradigm.
Section~\ref{sec:frankenstein} introduces \textit{Frankenstein} as a diagnostic framework and maps its logic onto recognition.
Section~\ref{sec:epistemicide} operationalizes the thesis by analyzing the pipeline as computational epistemicide.
Section~\ref{sec:metric-politics} models similarity as a norm.
Sections~\ref{sec:reform-fails} and \ref{sec:refusal} develop the critique of reform and the epistemology of refusal.

\section{Background: From Face to Vector}
\label{sec:from-face-to-vector}
\subsection{Pipeline overview}
\label{sec:pipeline-overview}
In contemporary systems, FRT can be summarized as a pipeline: capture $\rightarrow$ detection/alignment $\rightarrow$ embedding $\rightarrow$ comparison $\rightarrow$ decision. Deep learning systems popularized this logic as a practical stack for verification, often described as ``detect $\rightarrow$ align $\rightarrow$ represent $\rightarrow$ classify'' \cite{taigman2014deepface}. Contemporary systems increasingly treat recognition as metric learning in an embedding space \cite{schroff2015facenet,deng2019arcface}.

\begin{figure}[t]
\centering
\begin{tikzpicture}[
  stage/.style={draw, rounded corners=3pt, minimum width=9cm, minimum height=0.7cm, align=center, font=\small},
  arrow/.style={-{Latex[length=2.5mm]}, thick},
  info/.style={font=\scriptsize, align=left, text width=4cm},
  lost/.style={font=\scriptsize\itshape, align=right, text width=4cm, color=red!70!black}
]

\node[stage, fill=gray!5] (cap) at (0,0) {\textbf{1. Capture} (surveillance camera)};
\node[info] at (-3.2,-0.55) {\textcolor{blue!70!black}{${\sim}$1920$\times$1080 px (${\sim}$2M values)}};
\node[lost] at (3.2,-0.55) {scene, body, co-presence};

\draw[arrow] (0,-0.85) -- (0,-1.15);

\node[stage, fill=gray!10] (det) at (0,-1.5) {\textbf{2. Detect + Landmark} (RetinaFace: 5 points)};
\node[info] at (-3.2,-2.05) {\textcolor{blue!70!black}{5 landmarks (10 floats)}};
\node[lost] at (3.2,-2.05) {gesture, micro-expression};

\draw[arrow] (0,-2.35) -- (0,-2.65);

\node[stage, fill=gray!15] (ali) at (0,-3.0) {\textbf{3. Align + Warp} (affine to canonical pose)};
\node[info] at (-3.2,-3.55) {\textcolor{blue!70!black}{112$\times$112 px (12,544 values)}};
\node[lost] at (3.2,-3.55) {pose, illumination, non-canonical geometry};

\draw[arrow] (0,-3.85) -- (0,-4.15);

\node[stage, fill=blue!8] (emb) at (0,-4.5) {\textbf{4. Embed} (ResNet-100 $\to$ 512-dim unit vector)};
\node[info] at (-3.2,-5.05) {\textcolor{blue!70!black}{512 floats (ratio 4000:1 from step 3)}};
\node[lost] at (3.2,-5.05) {all structure beyond learned features};

\draw[arrow] (0,-5.35) -- (0,-5.65);

\node[stage, fill=blue!15] (dec) at (0,-6.0) {\textbf{5. Compare + Decide} (cosine sim $>$ threshold?)};
\node[info] at (-3.2,-6.55) {\textcolor{blue!70!black}{1 scalar $\to$ binary decision}};
\node[lost] at (3.2,-6.55) {continuous proximity $\to$ match/no-match};

\node[font=\scriptsize\bfseries, color=blue!70!black] at (-3.2,0.4) {Retained};
\node[font=\scriptsize\bfseries, color=red!70!black] at (3.2,0.4) {Destroyed};

\end{tikzpicture}
\caption{\rev{The FRT pipeline as progressive reduction. From ${\sim}$2 million pixel values to a single binary decision, each stage retains what serves comparison and destroys what exceeds the representational grammar.}}
\Description{Vertical pipeline showing five stages of face recognition with concrete dimensions retained at each step and information destroyed.}
\end{figure}
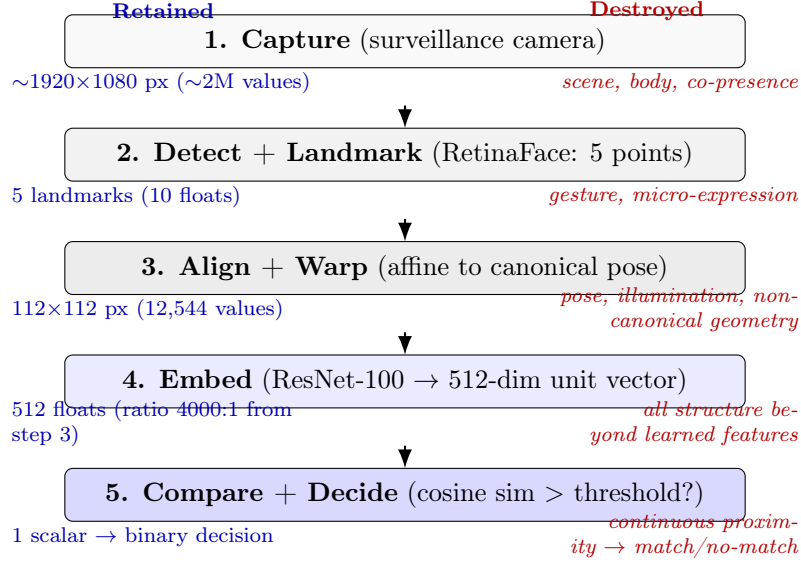

\subsection{Embeddings and metric learning}
\label{sec:embeddings-metric-learning}
A core shift in deep facial recognition is the transition from classification outputs to compact embeddings, in which distances encode identity similarity. FaceNet uses triplet loss to enforce a margin between \emph{same-person} and \emph{different-person} pairs in Euclidean space \cite{schroff2015facenet}. ArcFace enforces additive angular margins on normalized embeddings, increasing separability between identity clusters \cite{deng2019arcface}.

\subsubsection{Modern variants and the invariance of vectorization}
\label{sec:modern-variants}

While specific components have evolved rapidly (from stronger detectors and landmarkers to 3D-aware alignment and margin-based losses), the overall architecture of recognition remains remarkably stable: faces are transformed into fixed-dimensional embeddings, compared in a learned geometry, and adjudicated by decision rules. Modern face detectors (e.g., RetinaFace) improve localization under pose, blur, and occlusion \cite{deng2020retinaface}, and modern recognition objectives (e.g., ArcFace) sharpen separability by imposing geometric constraints on the embedding space \cite{deng2019arcface}. These upgrades change where error is concentrated and how robust systems appear, but they do not alter the epistemic premise of recognition: identity is still operationalized as proximity in a vector space \cite{schroff2015facenet}.

\rev{Figure~1 details this cascade for a contemporary ArcFace-based system. A surveillance scene ($\sim$2 million pixel values) is progressively reduced through detection (5 landmarks), canonical alignment (12,544 pixels), and embedding (512 floats) to a single cosine similarity score. The compression ratio from aligned face to vector alone is approximately 4,000:1. At each stage, what is retained serves comparison; what is destroyed is everything that exceeds the representational grammar. Every subsequent section analyzes what this reduction does, what it destroys, and what it authorizes.}

\section{Frankenstein as a Diagnostic Framework}
\label{sec:frankenstein}

Critical accounts of facial recognition often invoke literary metaphors to describe unintended consequences or technological hubris. This paper uses \textit{Frankenstein} differently: as a diagnostic framework for method. This section establishes the diagnostic conceptually; its technical operationalization across the pipeline is developed in Section~\ref{sec:epistemicide}, and its normative consequences for reform and refusal are developed in Sections~\ref{sec:reform-fails} and \ref{sec:refusal}. The novel foregrounds an epistemic procedure of disassembly and reconstruction, a way of producing a being through parts, and the ethical failure that follows when a life is made through a method that cannot care for what it produces \cite{shelley1818frankenstein}. The point is not that the creator ``misused'' a neutral technique; it is that the technique itself constitutes the creature.

\rev{What makes this procedure epistemic, rather than merely technical, is that the creature's mode of production determines what counts as knowledge about it. Victor Frankenstein does not encounter a living being and then describe it; he assembles parts into a whole and treats the resulting artifact as a legitimate representation of life. The creature's body is a theory: it embodies a claim that life can be known by dismantling it into components and stitching them into a standardized form. The ethical failure follows from this epistemic premise. Because the method of production defines what the creature is, and because that method cannot account for what was destroyed in the dismantling, the creature enters the world already dispossessed, knowable only on the terms set by the procedure that made it.}

A clarification on the scope of this epistemic claim is warranted. The argument is not that faces become generally unknowable, nor that human relational knowledge of faces ceases to exist; people continue to recognize one another in everyday life through forms of attention, memory, and context that no pipeline captures. The claim is that within the computational-institutional regime, the proxy is treated as the authoritative site of identity, and other forms of knowing the face are subordinated to it. What is foreclosed is not human recognition as such, but the epistemic standing of human recognition relative to the institutional decision.

\rev{Existing critical deployments of Shelley's novel in debates on artificial intelligence have typically treated it as a cautionary metaphor: technology may escape its creator's control, producing unintended consequences that demand ethical responsibility \cite{botting2020artificial}. The diagnostic proposed here is different in kind. It targets not the aftermath of creation but the method of creation itself: disassembly into measurable parts and reassembly into a proxy whose legitimacy is asserted by the procedure that made it. The Frankensteinian failure, on this reading, is not that the creator loses control but that the method of production is constitutively unable to care for what it produces, because what it produces was made by destroying what came before.}

\rev{To make the diagnostic operational, I propose three mappings. The \textbf{creator} is the institutional pipeline: the specific sociotechnical apparatus that decides to disassemble the face for identification and governance. The \textbf{creature} is the embedding: a numerical artifact assembled from the dissected face, designed to circulate across databases as a proxy for identity. \textbf{Who bears the consequences} is the person whose face was dissected to produce an artifact that now governs them. The Frankensteinian failure is structural: the pipeline has no channel of accountability between the proxy and the subject. This diagnostic names a deficit that ``bias'' cannot reach: the problem is not that the proxy is inaccurate but that the method of proxy-production is constitutively incapable of accounting for the subject it displaces. Table~\ref{tab:diagnostic} summarizes these mappings.}

\begin{table}[t]
\centering
\caption{\rev{The Frankenstein diagnostic: from novel to pipeline to generalized AI.}}
\label{tab:diagnostic}
\small
\begin{tabular}{p{2.5cm} p{4cm} p{4.2cm}}
\toprule
\textbf{Frankenstein} & \textbf{FRT Pipeline} & \textbf{AI (generalized)} \\
\midrule
\rev{Raw material} & \rev{The living face in context} & \rev{Any subject prior to datafication} \\
\addlinespace
\rev{Disassembly} & \rev{Detection, landmarking, alignment} & \rev{Feature extraction} \\
\addlinespace
Reassembly & Embedding: face compressed into vector & Model output: risk score, profile, label \\
\addlinespace
\rev{The creature} & \rev{The embedding: circulates across databases} & \rev{The proxy: circulates with institutional authority} \\
\addlinespace
\rev{The creator} & \rev{The institutional pipeline} & \rev{The deploying institution} \\
\addlinespace
\rev{Who suffers} & \rev{The person governed by the proxy} & \rev{The datafied subject} \\
\addlinespace
\rev{Ethical failure} & \rev{No accountability channel; ``the computer says it's you''} & \rev{No mechanism to contest the terms of conversion} \\
\addlinespace
\rev{Refusal} & \rev{Abolition of vectorized identity} & \rev{Refuse governance through computational proxies} \\
\bottomrule
\end{tabular}
\end{table}

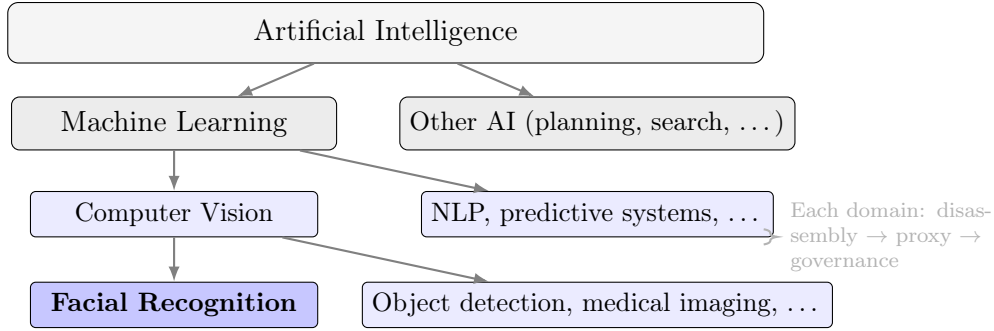
\begin{figure}[t]
\centering
\begin{tikzpicture}[
  bigbox/.style={draw, rounded corners=4pt, minimum width=10cm, minimum height=0.8cm, align=center, fill=gray!8},
  midbox/.style={draw, rounded corners=3pt, minimum width=4.3cm, minimum height=0.7cm, align=center, fill=gray!15},
  smallbox/.style={draw, rounded corners=2pt, minimum width=3.8cm, minimum height=0.6cm, align=center, fill=blue!8, font=\small},
  tinybox/.style={draw, rounded corners=2pt, minimum width=3.8cm, minimum height=0.6cm, align=center, fill=blue!22, font=\small\bfseries},
  arrow/.style={-{Latex[length=2mm]}, thick, gray}
]
\node[bigbox] (ai) at (0,0) {Artificial Intelligence};
\node[midbox] (ml) at (-2.8,-1.2) {Machine Learning};
\node[midbox] (other) at (2.8,-1.2) {\small Other AI (planning, search, \ldots)};
\node[smallbox] (cv) at (-2.8,-2.4) {Computer Vision};
\node[smallbox] (nlp) at (2.8,-2.4) {NLP, predictive systems, \ldots};
\node[tinybox] (frt) at (-2.8,-3.6) {Facial Recognition};
\node[smallbox] (othercv) at (2.8,-3.6) {Object detection, medical imaging, \ldots};
\draw[arrow] (ai) -- (ml);
\draw[arrow] (ai) -- (other);
\draw[arrow] (ml) -- (cv);
\draw[arrow] (ml) -- (nlp);
\draw[arrow] (cv) -- (frt);
\draw[arrow] (cv) -- (othercv);
\draw[decorate, decoration={brace, amplitude=5pt}, thick, gray!60]
  (5,-2.6) -- (5,-2.8) node[midway, right=6pt, font=\scriptsize, text width= 3cm, align=left]
  {Each domain: disassembly $\to$ proxy $\to$ governance};
\end{tikzpicture}
\caption{\rev{FRT within the broader field of AI. The Frankenstein diagnostic targets the epistemic operation shared across all systems that convert human subjects into computable proxies.}}
\label{fig:ai-field}
\Description{Hierarchical diagram: AI, Machine Learning, Computer Vision, Facial Recognition highlighted.}
\end{figure}

\rev{A clarification of scope is necessary. This paper analyzes facial recognition as a case study, but the diagnostic is not confined to a single application (Figure~\ref{fig:ai-field}). Predictive policing, credit scoring, automated hiring, and natural language processing all perform variants of the same Frankensteinian operation. Facial recognition is exemplary because the pipeline makes the structure of disassembly unusually visible.}

\rev{A brief illustration demonstrates the diagnostic's portability. In predictive policing systems such as PredPol (now Geolitica), the ``raw material'' is not a face but a neighborhood's history of reported crime. The ``disassembly'' converts this history into spatial and temporal features. The ``creature'' is a risk score assigned to a geographic cell, designed to circulate across patrol schedules and resource allocation decisions. The ``creator'' is the institutional apparatus (police department, vendor, city administration) that decides to convert lived geography into governable data. Who bears the consequences are the residents of flagged areas, disproportionately Black and Brown, who are subjected to intensified surveillance on the basis of a proxy they did not authorize and cannot contest. The structural pattern is identical to FRT: disassembly, compression into proxy, institutional governance on the basis of the artifact, and the absence of any channel through which the governed subject can refuse the terms of the conversion.}

\rev{This visibility is sharpened by a genealogy that precedes AI. The statistical methods that underpin machine learning were developed in the service of eugenics: Galton coined the term in 1883 and invented regression to the mean \cite{galton1883inquiries}; Pearson developed the correlation coefficient as Galton Chair of Eugenics \cite{pearson1901national}; Fisher developed discriminant analysis while promoting eugenic policy \cite{fisher1930genetical}. MacKenzie \cite{mackenzie1981statistics} has shown these innovations were developed in direct service of eugenic ideology. Crawford \cite{crawford2021atlas} argues that AI systems are designed to classify, and classification is never neutral. Chun \cite{chun2021discriminating} sharpens the point: correlation stems from eugenic attempts to ``breed'' a better future, and default assumptions encode segregation even without race as an explicit variable. From Galton's composite photographs to contemporary embedding spaces, the method is one of disassembling human particularity into a standardized representation and authorizing the artifact as truth. What changes is the medium and the scale, not the epistemic operation.}

Read through this diagnostic, embedding-based recognition appears as a modern procedure of dismemberment and stitching: the face is isolated, abstracted, forced into canonical geometry, and reanimated as a proxy. \rev{This methodological constant has a genealogy. Sekula \cite{sekula1986body} showed that nineteenth-century police photography constructed the ``criminal body'' through standardized measurements and bureaucratic archives; Gandy \cite{gandy1993panoptic} generalized this as the ``panoptic sort.'' The face recognition pipeline inherits this logic, operating through metric geometry rather than paperwork.}

\rev{Stark and Hutson \cite{stark2022physiognomic} and Aguera y Arcas et al.\ \cite{aguerayarcas2017physiognomy} define ``physiognomic AI'' as inferring character or social outcomes from physical characteristics. This paper locates the harm at a prior stage: in the vectorization itself, before any physiognomic inference.} This method participates in a broader onto-epistemic order in which recognizability is organized by raciality \cite{silva2007toward}. Carneiro \cite{carneiro2005construcao} names epistemicide as the production of ``non-being'' through the denial of epistemic authority; Browne \cite{browne2015dark} shows how surveillance renders Blackness as an object of capture; Benjamin \cite{benjamin2019race} argues that contemporary innovations repackage racial governance as neutrality. In this lineage, \textit{Frankenstein} is a diagnostic for a method that assembles identity by dissection and authorizes that assembly as truth.

\section{Computational Epistemicide: From Face to Proxy}
\label{sec:epistemicide}

In this section, I name the operational logic of contemporary face recognition as \emph{computational epistemicide}: a process by which the face is stripped of its relational, historical, and situated meanings so it can be rendered legible within a narrow computational grammar. The epistemic violence is not an accident of biased data; it is enacted by the pipeline's requirement that the face be transformed into a governable proxy. Building on the diagnostic introduced in Section~\ref{sec:frankenstein}, I treat each stage of the pipeline as an epistemic operation: what it technically does, what it discards, \rev{why the discard constitutes an epistemic reduction, and in which institutional contexts this reduction becomes consequential.}

\rev{\subsection{Canonicalization and the inseparability of norms}}

\rev{Before tracing the pipeline stage by stage, a foundational clarification is required. One might ask: is canonicalization harmful because it encodes specific (racialized) norms (that is, a ``white face''), or is it harmful as such, regardless of which norms shape the canonical template? My argument is that these positions are not separable, and collapsing them into an either/or obscures the structure of the harm. Drawing on Ferreira da Silva's \cite{silva2007toward} theorization of raciality as a global onto-epistemic order, I claim that there is no abstract position from which canonicalization can be performed. The pipeline does not encode ``a white face'' as a single demographic template; it enforces a regime of standardization that inherits historical hierarchies of who has counted as legible and whose appearance has been treated as deviation. Canonicalization is not an abstract epistemic wrong independent of history; if it were, any measurement (a thermometer, a scale) would be violence. Nor is the harm reducible to a demographic bias that better training data could correct; if it were, the reformist position would be sufficient. The harm is that canonicalization is an operation that can only be performed from within an onto-epistemic order that has already distributed recognizability unevenly, and the pipeline has no mechanism for contesting or revising that distribution.}

\paragraph{Selection: detection and cropping.}
Face detection and cropping are often described as upstream necessities: isolate the face region and remove background clutter. Operationally, however, detection is a selection mechanism that defines the face as a bounded object extractable from context \cite{viola2001rapid,deng2020retinaface}. The crop enacts an epistemic boundary: it retains what the detector can stabilize as ``face'' and removes the surrounding relational scene (\rev{body, gesture, environment, co-presence}) as irrelevant by design. \rev{The epistemic character of this operation lies in what it decides in advance: that the face is a self-contained object whose meaning does not depend on context. In governance contexts (law enforcement, border control, welfare verification), the crop is the first step by which a person is converted into a subject of computational adjudication. A face captured at a checkpoint, extracted from a protest crowd, or isolated from a welfare applicant's body is not merely ``detected''; it is conscripted into a pipeline whose downstream operations will convert it into a categorical decision.}

\paragraph{Abstraction: landmarking as anatomical coordinates.}
Landmarking translates the face into fiducial points that serve as anchors for subsequent normalization. This step performs an anatomical abstraction: the face becomes what can be expressed as a coordinate system. What is preserved are points and proportions; what is discarded are temporal expressivity, micro-gesture, and situated meaning. \rev{The loss is concrete. The same person may appear radically different across temporal scales: at the micro-scale, makeup, mood, and lighting alter facial appearance moment to moment; at the meso-scale, health, fatigue, and emotional state reshape facial geometry over days and weeks; at the macro-scale, aging, injury, and medical treatment transform facial structure over years. Identity, as lived, as recognized in social interaction, accommodates these variations through context, memory, and relational knowledge. Landmarking freezes the face at a single moment and reduces its expressive range to a coordinate configuration, treating all variation that exceeds the landmark set as noise.} Errors at this stage propagate forward as degraded normalization and representation (see the pipeline framing in Section~\ref{sec:pipeline-overview} and embedding sensitivity discussed in \cite{taigman2014deepface,schroff2015facenet}).

\paragraph{Normalization: warping/alignment and the production of a canonical geometry.}
Alignment and frontalization are presented as ``corrections'' that remove nuisance variation so faces become comparable. Yet warping a face into a standardized pose is not merely technical; it enforces a canonical geometry as the condition of recognizability. In modern embedding pipelines, alignment is treated as a prerequisite for stable representation learning \cite{taigman2014deepface}.

\paragraph{The canonical face and the production of a form-subject.}
Alignment and frontalization do more than ``standardize'' input; they produce what I call the \emph{canonical face}: a normative template that determines which facial variations count as legitimate signal and which are treated as distortion. The canonical face is not a demographic average; it is a geometric and photometric discipline imposed as the precondition for comparability. \rev{As Dumit \cite{dumit2004picturing} demonstrated for brain imaging, the scientific production of a ``normal'' template does not describe the world but constitutes it, embedding assumptions into every stage of the pipeline. The PET scan's ``normal brain'' and the face recognition pipeline's ``canonical face'' operate analogously: both are constructed standards that present themselves as objective descriptions. Empirically, the canonical face is not a single demographic template but an operational effect: alignment enforces a standardized geometry, and metric learning produces regions of stability and fragility that are not uniformly distributed across demographic groups. NIST FRVT evaluations documented elevated false positive rates for women of African descent relative to Eastern European men in one-to-many searches \cite{grother2019frvt}, indicating that the geometry of the embedding space encodes asymmetric legibility.} Once recognizability depends on conformity to this template, the pipeline implicitly produces a corresponding \emph{form-subject}: an identity that can appear as a stable point in space only by undergoing canonicalization. What cannot be stabilized under this discipline is not merely hard to recognize; it is positioned as epistemically excess: noise, nuisance variation, or misalignment to be repaired.

\paragraph{Compression: embedding as fixed-dimensional proxy.}
Embedding completes the substitution by compressing the canonicalized face into a fixed-dimensional vector intended to stand in for identity \cite{taigman2014deepface,schroff2015facenet}. Metric-learning objectives explicitly engineer a space in which within-identity distances shrink and between-identity distances grow, turning ``similarity'' into a computable property of distance or angular proximity \cite{schroff2015facenet,deng2019arcface}. The result is not a richer description of the face but a proxy optimized for comparison, a numerical artifact that can circulate through databases and institutions precisely because it has been stripped of relational context. \rev{The institutional consequence is direct. In law enforcement, the threshold converts metric proximity into suspicion, with false positives translating into coercion \cite{garvie2016perpetual}. In border control, the vectorized face is compared against watchlists, and errors translate into detention or refusal of entry. In welfare verification, the canonical face determines who can access services; failure to ``configure'' as a legible face means exclusion from rights \cite{eubanks2018automating}. In each context, the costs of error fall disproportionately on those whose faces are least well served by canonicalization.}

\rev{These consequences are documented. In the United States, at least eight persons have been wrongfully arrested on the basis of false facial recognition matches, all of them Black \cite{garvie2016perpetual}. Robert Williams was detained for thirty hours in Detroit; Porcha Woodruff, eight months pregnant, was held for eleven hours; Alonzo Sawyer spent days in jail for an assault he did not commit. In each case, the proxy governed: the vector, not the person, was the basis of the institutional decision. In Brazil, 90.5\% of persons arrested through facial recognition between March and October 2019 were Black; in Rio de Janeiro, 81\% of unjust arrests based on photographic recognition involved Black individuals. In India, the Aadhaar system conditions access to welfare on biometric legibility. These cases are not artifacts of poor implementation; they are structural consequences of a pipeline that converts faces into proxies and governs persons on the basis of the conversion.}

\paragraph{Computational epistemicide as a regime of authorized identity.}
To name this sequence as \emph{computational epistemicide} is not to claim that computation is merely ``wrong'' about faces. It is to claim that the pipeline installs a regime of knowledge in which the face can be recognized only insofar as it can be reduced to what the system can stabilize as data. This is why the violence is not exhausted by disparate error rates. Even when a system ``works'' as designed, it works by declaring that the authorized site of identity is the proxy, and by subordinating the living face to the vector that stands in for it.

\rev{This is also why requiring institutional legibility is not normatively neutral. The wrong is not identification as such; it is that canonicalization-based identification imposes a specific representational grammar as the condition of recognition, under asymmetric stakes. The person who must submit their face to canonicalization in order to access welfare, cross a border, or avoid suspicion does not choose the terms of the grammar. When those terms are derived from training distributions that reflect historical hierarchies of recognizability, compulsory legibility becomes a mechanism for reproducing those hierarchies under the appearance of technical neutrality.}

My use of epistemicide extends the concept as developed by Sueli Carneiro: the production of the Black subject as ``non-being'' through systematic forms of erasure, negation, and the denial of epistemic authority \cite{carneiro2005construcao}. In face recognition, erasure is operationalized: what exceeds the representational grammar (relational context, situated meaning, and forms of appearance not well served by canonicalization) is converted into nuisance variation to be corrected, discarded, or ignored. The pipeline thus does not merely \emph{fail} to know the face; it defines in advance what counts as knowable, and it enforces that definition through preprocessing, geometry, and metric comparison. Situated within raciality as a global onto-epistemic order \cite{silva2007toward}, computational epistemicide can be read as a contemporary technique for producing legibility by imposing a norm of recognizability.

\rev{A potential objection must be addressed. If surveillance renders Black subjects hypervisible (Benjamin \cite{benjamin2019race}), how can the pipeline be described as epistemicide? The answer is that computational epistemicide and hypervisibility are structurally linked, not opposed. The pipeline renders the subject visible only as what the system can stabilize as data. In Carneiro's formulation, epistemicide is not literal invisibility but the denial of epistemic authority: the foreclosure of the subject's capacity to define the terms under which they are known. The pipeline operationalizes this denial by recognizing the face only insofar as it submits to canonicalization. What is destroyed is not visibility but epistemic sovereignty over one's own appearance.}

\rev{This extension must be situated within its intellectual genealogy. Santos \cite{santos2014epistemologies} formulated epistemicide as the destruction of knowledge systems through colonial imposition. Carneiro \cite{carneiro2005construcao} shifted the register to subject-formation: epistemicide as the production of ``non-being'' through the denial of epistemic authority. The computational extension does not depart from Carneiro; it operationalizes her insight. In Carneiro's formulation, non-being is the foreclosure of the subject's capacity to define the terms under which they are known. The pipeline performs precisely this: it defines in advance what counts as a face, similarity, and identity, and the subject has no mechanism to contest those definitions. The extension specifies the mechanism with a precision that philosophical formulation alone cannot provide: the architectural decisions of the pipeline (loss function, embedding dimensionality, comparison geometry, threshold logic) perform the ontological displacement. This is why the analysis requires a vantage point within computational science: these decisions are legible as decisions, rather than as neutral technical necessities, only to those who understand how pipelines are designed.}

This framing clarifies a central implication for normative analysis: facial recognition is not best understood as a descriptive technology that sometimes errs, but as a constitutive technology that decides what a face must be in order to count as an identity. The next section makes this explicit by analyzing similarity-as-distance and thresholding as normative operators: they turn the proxy's geometry into an institutional criterion of ``close enough.''

\section{Distance, Similarity, and Eugenic Normality}
\label{sec:metric-politics}

This section does not evaluate metrics; it models what it means, epistemically and normatively, to define identity through an embedding geometry. The argument targets the normative role of distance as such: it is not a neutral translation of faces into numbers but a normative apparatus that makes ``similarity'' into a criterion of legibility and converts technical comparability into institutional authority.

\paragraph{Similarity is not discovered; it is stipulated.}
In embedding-based recognition, the question ``Are these the same person?'' is rewritten as ``Are these points close enough?'' That rewrite is an epistemic decision. It replaces the open-ended, contextual, and contested character of identity with a single formal relation: distance. Once similarity is defined as a metric relation, the system no longer asks what a face means; it asks whether a proxy satisfies a geometric constraint. This is why ``similarity'' must be treated as a norm: it is not a property of faces in the world, but a criterion imposed by a representational scheme.

\rev{The problem with this formalization is not that it is approximate but that it is constitutive: it replaces one kind of question with a categorically different one. ``Is this the same person?'' admits of degrees, contexts, and negotiations. ``Are these points close enough?'' admits of exactly one answer for a given threshold, context-free and non-negotiable. The formalization abolishes the first question and substitutes a different one. What is lost is the ontological character of the relation: identity ceases to be something recognized between subjects and becomes something computed between objects.}

\paragraph{Optimization sculpts centers of density and margins of uncertainty.}
Because the embedding space is learned under particular conditions of capture, labeling, and preprocessing, it does not distribute faces uniformly. It produces regions of stability and regions of fragility: centers where the model's notion of closeness is tight and reliable, and margins where proximity becomes ambiguous. This is not a statistical footnote; it is a structural effect of making recognition depend on standardization. When upstream steps (selection, abstraction, canonicalization) succeed, the proxy tends to land in regions where distances behave as intended. When those steps fail or degrade, the proxy tends to drift toward margins where ``close enough'' becomes harder to interpret and more dependent on institutional tolerance for error. In this way, the geometry of the space becomes an epistemic map of who is easily legible and who is persistently uncertain.

\paragraph{Thresholds operationalize ``close enough'' as governance.}
The decision boundary turns that map into power. A threshold converts a continuous distance into a categorical outcome: match or non-match; access or denial; suspicion or clearance. Importantly, the threshold does not merely reflect similarity; it enacts it. It states, institutionally, how much deviation will be tolerated before identity is refused. When recognition is used in governance contexts, this tolerance is never purely technical: it is a policy about which errors are acceptable, for whom, and under what consequences. Thresholding is therefore the point where metric geometry becomes a norm with teeth.

Taken together, these points clarify the core claim of this paper: once identity is defined as distance in an embedding geometry, recognition becomes inseparable from standardization. The metric is not just a measurement tool inside a pipeline; it is the mechanism by which a canonical notion of personhood is imposed, by which uncertainty is distributed as marginality, and by which institutional decisions are naturalized as ``similarity.''

\paragraph{From geometry to institutional authority.}
Embeddings do not remain inside a model. They circulate as institutional objects: stored, indexed, shared across systems, and mobilized as evidence in workflows that demand rapid decisions. The ``decision boundary'' is not merely a statistical threshold; it is an organizational instrument that translates model geometry into operational authority. The embedding and similarity score appear as technical facts, while the choice of threshold encodes institutional tolerance for error under asymmetric stakes: false positives become suspicion and coercion; false negatives become denial of access. Because these costs are not borne equally, the system's promise of objectivity functions as depoliticization: it naturalizes the conversion of facial comparability into social truth while obscuring the power relations that make the cut consequential.

\section{Beyond Bias: Why ``Ethical AI'' Optimization Fails}
\label{sec:reform-fails}

In the dominant ``ethical AI'' imagination, the harms of facial recognition are framed as failures of execution: insufficiently diverse datasets, miscalibrated thresholds, poor documentation, or inadequate evaluation \cite{gebru2018datasheets,mitchell2019modelcards,raji2020accountabilitygap}. This reformist grammar presumes that recognition is, in principle, a legitimate task, and that justice can be achieved by improving performance and governance around the system \cite{dwork2012awareness,hardt2016eopp,kleinberg2017tradeoffs,chouldechova2016fairprediction}. I argue instead that the core harm is ontological: facial recognition does not merely misrecognize; it produces a regime of recognizability by reducing the face to a proxy and making identity a metric relation in a learned space (Sections~\ref{sec:epistemicide}--\ref{sec:metric-politics}). Under this premise, optimization cannot ``fix'' the violence because the violence is enacted by the representational method itself.

Empirically, disparities in classification and matching accuracy are real and measurable, and they have been documented across commercial systems and benchmarking regimes \cite{buolamwini2018gender,grother2019frvt}. Yet treating these disparities as the primary problem risks mistaking symptoms for structure. Even if performance were equalized across demographic categories, the system would still require canonicalization, compression, and thresholding to convert faces into governable identities. In other words, ``fairer'' recognition would still be recognition: a pipeline that extracts faces from their contexts, normalizes them into a canonical geometry, and adjudicates personhood through a technical cut. More broadly, the reformist impulse often relies on abstraction that strips systems from the institutional contexts that produce and deploy them, turning political questions into parameter choices \cite{selbst2019abstraction}.

This ontological critique becomes sharper when situated within the global production of race. A system that defines identity through comparability and standardization does not operate on neutral human variation; it operates within an onto-epistemic order in which raciality organizes who can be recognized as fully human and on what terms \cite{silva2007toward,carneiro2005construcao,wynter2003unsettling}. Facial recognition's promise of universal legibility is therefore inseparable from a historical project of making subjects governable through classification. \rev{Indeed, facial recognition enables a further transformation that sharpens this critique. Because the system operates on individual vectors rather than categorical labels, it allows the exercise of power on a face-by-face basis, no longer requiring race as a visible sorting criterion. Computation can now classify individual by individual, producing the hierarchies that racial categories once organized, but through metric proximity rather than phenotypic classification. The system does not need to encode race as an explicit variable to reproduce racialized outcomes; the geometry of the embedding space, shaped by the historical conditions of its production, does this work without naming it. In this sense, FRT does not merely inherit racial governance; it refines it, dispensing with the concept of race as a crude rubric while intensifying and individualizing its effects.} Surveillance infrastructures have long rendered Blackness as an object of capture and management, not as a subject of mutual recognition \cite{browne2015dark,noble2018algorithms}. In this lineage, ``ethical'' optimization risks becoming a new language for the same old settlement: improving the apparatus that decides who is legible, rather than refusing the apparatus's authority to decide \cite{benjamin2019race}.

Finally, reform often functions politically as legitimacy. When institutions can claim that systems are ``more accurate'' or ``less biased,'' they gain justification to expand deployment into domains where contestation is difficult and the consequences are coercive. In this sense, improvement can become entrenchment: it strengthens the apparatus's credibility without confronting whether the apparatus should exist.

\rev{A counter-argument must be confronted directly. Proponents contend that facial recognition serves urgent public goods: locating missing children, identifying suspects in violent crimes, preventing terrorist attacks. These cases are real. But the argument from exceptional benefit performs a specific rhetorical operation: it presents the extraordinary case as justification for a permanent infrastructure. The child found through a database match is invoked to legitimize a pipeline that operates continuously on entire populations, most of whom are not missing, not suspected, and not consenting.}

\rev{The question is not whether identification can sometimes produce good outcomes; it is who builds the infrastructure, who decides which faces populate the database, who sets the threshold, and who bears the consequences when the system errs. In every documented deployment, the populations most subjected to surveillance are not the populations that decide whether and how the technology is deployed. The securitarian justification obscures an asymmetry of power that is constitutive, not incidental: the infrastructure is built by those who will not be governed by it, and imposed on those who have no mechanism to contest it. The abolitionist stance advanced here is therefore not a demand for better recognition or fairer surveillance. It is a refusal of the premise that any institution should possess the authority to convert human faces into governable data points and to treat proximity in a vector space as a legitimate basis for decisions about freedom, access, and suspicion.}

If facial recognition is structurally oriented toward producing a canonical face and governing by similarity, then ethical work cannot be limited to tuning the proxy. It must include the refusal of vectorized identity as an acceptable basis for rights, access, and belonging.

\section{Refusal as Consequence: Abolition Beyond Optimization}
\label{sec:refusal}

If the harm is ontological, the appropriate response is not a better proxy but a refusal of proxyhood. I use ``abolition'' here not as an incremental toolkit inside the system, but as a normative and epistemological stance: the rejection of the vector as a valid substitute for personhood and the rejection of the institutional impulse to make facial legibility the condition of participation \cite{benjamin2019race,wynter2003unsettling}. This stance follows from the preceding diagnosis: if the face must be dismantled to become computable, then the resulting computation cannot be treated as a neutral instrument for justice (Sections~\ref{sec:epistemicide}--\ref{sec:metric-politics}).

Frankenstein offers a diagnostic vocabulary for this refusal. The modern Prometheus does not merely create; he dismembers, assembles, and then demands that the creature bear the consequences of a method that cannot care for what it produces \cite{shelley1818frankenstein}. In facial recognition, the ``creature'' is the embedding: a fixed-dimensional artifact designed to circulate across databases and institutions. The ethical failure is not an accident of bias but a consequence of a methodology that requires dissection to function. Refusal, then, is not a refusal of technology in the abstract; it is a refusal of a specific epistemic operation: the conversion of the face into a governable geometry and a metric cut.

\rev{The ethical core of Shelley's novel illuminates why refusal, rather than reform, is the appropriate response. Victor Frankenstein does not fail because he creates; he fails because, having created through dissection and reassembly, he refuses to respond to what he has produced. The creature demands recognition, relation, accountability. Victor refuses. The violence that follows is the consequence of a method that severs every channel of care between maker and made.}

\rev{In the face recognition pipeline, the dispossession is deeper. The ``creature,'' the embedding, cannot speak at all; it is a numerical artifact with no capacity to contest its own terms. And the person whose face was dissected has no channel within the system through which to say ``this is not me.'' Robert Williams said exactly this to the Detroit police officer who showed him the surveillance image. The officer replied: ``The computer says it's you.'' The system had no mechanism to hear the refusal. This is the Frankensteinian failure transposed to computation: not that the proxy is wrong, but that the infrastructure of proxy-production has no organ of listening, no channel of accountability, no possibility of care.}

This refusal is grounded in Black feminist and decolonial epistemologies that treat knowledge not as a neutral description but as a terrain of struggle over who can be constituted as a subject \cite{figueiredo2020epistemologia,silva2007toward,carneiro2005construcao}. It also resonates with the claim that modern regimes of power secure themselves by overrepresenting a particular genre of the Human, rendering other ways of being as derivative, deviant, or unrecognizable \cite{wynter2003unsettling}. From this perspective, the problem is not that some faces are misread, but that the apparatus claims the authority to read faces at all, to translate them into data and to naturalize that translation as identity. Abolition names the disruption of that authority.

Refusal also names a right to opacity: a refusal to be made transparent and commensurable under a dominant representational grammar \cite{glissant1997poetics}. In the context of facial recognition, opacity is not an aesthetic preference; it is a political demand against compulsory legibility. Accounts of racial governance show how surveillance and classification secure power by defining the terms of visibility and legibility, and by turning Blackness into an object of capture and management rather than a subject of mutual recognition \cite{browne2015dark,benjamin2019race}. Abolition, here, is the insistence that institutions do not get to set the conditions under which a face becomes a valid claim to personhood.

Abolition as refusal is also pedagogical: it reorients the question from ``How do we improve recognition?'' to ``Why must recognition be the solution?'' \cite{freire1996pedagogy,tuckyang2014unbecoming}. It insists that dignity and access cannot depend on compliance with a canonical template, and that rights cannot be conditional on the production of a metric self.

\rev{Concretely, refusal opens space for institutional alternatives that do not require the computational reduction of human life to dissectible data points. At least three directions are available. First, \emph{self-sovereign identity} systems allow individuals to hold and selectively disclose verifiable credentials without submitting biometric data to centralized databases, placing control over identity with the subject rather than the institution. Second, \emph{community-based verification} practices, in which identity is attested through relational networks rather than algorithmic matching, offer models of recognition that are accountable to the communities they serve. Third, \emph{institutional redesign} that decouples access to rights and services from biometric identification altogether, replacing compulsory legibility with consent-based or presential verification, addresses the root demand: that no institution should require the dissection of the face as the price of participation.} The aim is not to perfect facial recognition, but to displace it as a legitimate instrument of governance.
\section{Conclusion}
\label{sec:conclusion}

This paper argued that embedding-based facial recognition enacts \emph{computational epistemicide}: it makes the face knowable only by dismantling it into a governable proxy. By treating detection/cropping, landmarking, alignment/frontalization, embedding, and thresholding as epistemic operations, I showed how the pipeline produces a canonical face as the condition of legibility and a corresponding form-subject as the condition of recognition. Under this premise, identity is operationalized as proximity in a learned geometry, and ``similarity'' becomes a normative criterion of who can be recognized as ``close enough.''

The implications are twofold. First, because the harm is enacted by the representational method itself, reformist ``ethical AI'' approaches that focus on optimizing accuracy, calibration, or documentation remain structurally insufficient: they refine the proxy without contesting the legitimacy of proxyhood. Second, if vectorized identity functions as an institutional instrument of governance, then the appropriate response is not to seek better embeddings but to adopt refusal as a normative stance: rejecting the vector as a legitimate basis for rights and access, and displacing facial recognition as a technology of authority.

\rev{A broader implication follows. The Frankenstein diagnostic identifies a structural pattern (disassembly, compression, proxy-governance) that recurs across AI systems operating on human subjects. If this diagnostic is correct, then the question ``How do we make recognition fairer?'' must be preceded by ``Should the computational reduction of the face be the basis of institutional decisions at all?'' This paper has argued from within computational science, not against it. The critique is not that computation is inherently violent, but that a specific class of operations, those that produce governable proxies from living subjects without mechanisms of accountability, enacts epistemic violence regardless of performance. Different critical lenses (sociological, juridical, philosophical, technical) illuminate different dimensions of this problem; they are mutually necessary, not in competition. What the technical lens uniquely contributes is the identification of the precise architectural decisions where the epistemic displacement is constituted. Abolition is the demand to dismantle the conditions that make computational dissection appear as recognition, and to build institutional alternatives that do not require the face to be destroyed in order to be known.}

\bibliographystyle{plainnat}
\bibliography{references}

\section*{Generative AI Usage Statement}
No generative AI tools were used in the production of this manuscript. All substantive claims, technical descriptions, theoretical framing, and final wording were authored by the human author.

\section*{Ethical Considerations}

This paper analyzes facial recognition as a regime of governability. The ethical stakes center not on whether systems can be made more accurate, but on whether it is legitimate to operationalize identity through vectorized proxies at all. The concept of computational epistemicide cautions against framing the problem as bias mitigation: interventions that improve performance without contesting the proxy's authority risk legitimizing and expanding the apparatus. This paper does not provide implementation guidance for deploying facial recognition systems. Instead, it argues that certain harms are produced by the representational method itself, and that responsible practice may require non-use, de-deployment, and institutional alternatives to biometric identification.

\end{document}